\documentclass[doublecol]{epl2} 
\usepackage{graphicx}
\usepackage{booktabs}
\usepackage{siunitx}

\usepackage{amsmath, amssymb, graphics, graphicx}
\usepackage{textcomp, subfigure}\usepackage{color}
\usepackage[final]{pdfpages}

\definecolor{linkcolor}{rgb}{0,0,0.6} 
\usepackage[pdftex,colorlinks=true, pdfstartview=FitV, linkcolor= linkcolor, citecolor= linkcolor, urlcolor= linkcolor, hyperindex=true,hyperfigures=true]{hyperref} 

\begin{document}
\title{Detailed Jarzynski Equality applied  to \\ a Logically Irreversible Procedure}
\author{A. B\'{e}rut, A. Petrosyan, S. Ciliberto}
\institute{Laboratoire de Physique ENS Lyon (CNRS UMR5672), 46, all\'{e}e d'Italie
F69007 Lyon, France}

\date{\today}

\begin{abstract}
{A single bit memory system is made with a brownian particle held by an  optical tweezer in a double-well potential and the work necessary to erase the memory is measured. We show that the minimum of this work  is close to  the Landauer's bound only for very slow erasure procedure. Instead a  detailed Jarzynski equality allows us to retrieve the Landauer's bound independently on the speed of this erasure procedure. For the two separated subprocesses, i.e.  the transition from state 1 to state 0 and the transition from state 0 to state 0, the Jarzynski equality does not hold but the generalized version links the work done on the system to the probability that it returns to its initial state under the time-reversed procedure.}
\end{abstract}


\maketitle

The connection between thermodynamics and information is nowadays a widely studied problem \cite{ref:Benett,ref:Vedral,ref:newsVandenBroeck,ref:Seifert,ref:reviewSeifert}. The main questions concern the amount of energy necessary in order to perform a logical operation in a given time  and how the information entropy is  related to the free energy difference between the initial and final state of this logical operation.  In this context the Landauer's principle \cite{ref:Landauer} is very important as it states that  for any irreversible logical operation the minimum amount of entropy production  is $-k_{\text{B}} \ln(2)$ per bit commuted by the logical operation, with $k_{\text{B}}$ the Boltzmann constant. Specifically a logically irreversible operation is an operation for which the knowledge of the output does not allow to retrieve the initial state, examples are logical AND, OR and erasure. In a recent paper \cite{ref:Nous} we have experimentally shown that indeed the mini
 mum amount of work necessary to erase a bit is  actually associated with this Landauer's bound which can be asymptotically reached for quasi-static transformations. The question that arises naturally is whether this work corresponds to the free energy difference between the initial and final state of the system. To answer to this question it seems natural to use the Jarzinsky equality \cite{ref:Jarzynski} which allows one to compute the free energy difference between two states of a system, in contact with a heat bath at temperature $T$. When such a system is driven from an equilibrium state A  to a state B through any continuous  procedure, the Jarzynski equality links the stochastic work $W_{\text{st}}$  received by the system during the procedure to the free energy difference $\Delta F = F_{B}-F_{A}$ between the two states:
\begin{equation}
\left\langle e^{- \beta W_{\text{st}}} \right\rangle = e^{-\beta \Delta F}
\label{eq_Jeq}
\end{equation}
\noindent Where  $\left\langle . \right\rangle$ denotes the ensemble average over all possible trajectories, and $\beta = \frac{1}{k_{\text{B}}T}$ (see eq. \ref{eq:stocwork} for the precise definition of the work $W_{\text{st}}$). 
 
In this letter we  analyze the question of the application of eq. \ref{eq_Jeq} for estimating the $\Delta F$ corresponding to the erasure operation in our experiment, in which a colloidal particle confined in a double well potential is used as a single bit memory. We will show that the classical Jarzynski equality (eq. \ref{eq_Jeq}) is not useful here but that a detailed Jarzynski Equality \cite{ref:Jarzynski2} allows us to retrieve the Landauer limit independently of the work done on the system during the memory erasure procedure, and to link this work to the probability that the system returns to its initial state under the time-reversed procedure.
\bigskip




The setup has already been described in a previous article \cite{ref:Nous} and we recall here only the main features. 

\begin{figure}[ht!]
\begin{center}
\includegraphics[width=8cm]{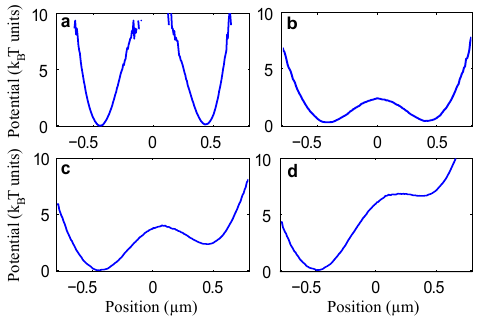}
\caption{Measurement of the potential in which the silica bead is confined, with no external force for two different laser powers (\textbf{a}: $\SI{48}{\milli\watt}$ and \textbf{b}: $\SI{15}{\milli\watt}$), and extrapolation for low power with two different value of the force (\textbf{c}: $\SI{10}{\femto\newton}$ and \textbf{d}: $\SI{30}{\femto\newton}$). The potential is computed from the equilibrium probability density function of the bead's position.} \label{fig:potential}
\end{center}
\end{figure}

A custom-built vertical optical tweezers is used to realize a two-state system: a silica bead (radius $R = \SI{1}{\micro\meter}$) is trapped at the focus of a laser beam (wavelength \SI{1024}{\nano\meter}) which is rapidly switched (at a rate of \SI{10}{kHz}) between two positions (separated by \SI{1.45}{\micro\meter}) using an acousto-optic deflector. A disk-shaped cell (\SI{18}{\milli\meter} in diameter, \SI{1}{\milli\meter} in depth) is filled with a solution of beads dispersed in bidistilled water at low concentration. The bead used for the experiment is trapped by the laser and moved into the center of the cell (with gap $\sim \SI{80}{\micro\meter}$) to avoid all interactions with other beads. The bead is trapped at $\SI{25}{\micro\meter}$ above the bottom surface of the cell, it feels a double-well potential with a central barrier varying from $2 k_{\text{B}}T$ to more than $8 k_{\text{B}}T$ depending on the power of the laser (see figure \ref{fig:potential}, \textbf{a}
  and \textbf{b}). The left well is called ``0'' and the right well ``1''. The position of the bead is tracked using a fast camera with a resolution of \SI{108}{\nano\meter} per pixel, which after treatment gives the position with a precision greater than \SI{10}{\nano\meter}. The trajectories of the bead are sampled at \SI{502}{\hertz}.

The logical operation performed by our experiment is the erasure procedure. This procedure brings the system initially in one unknown state (0 or 1 with same probability) in one chosen state (e.g. 0). It is done experimentally in the following way.

\begin{figure}[ht!]
\begin{center}
\includegraphics[width=8cm]{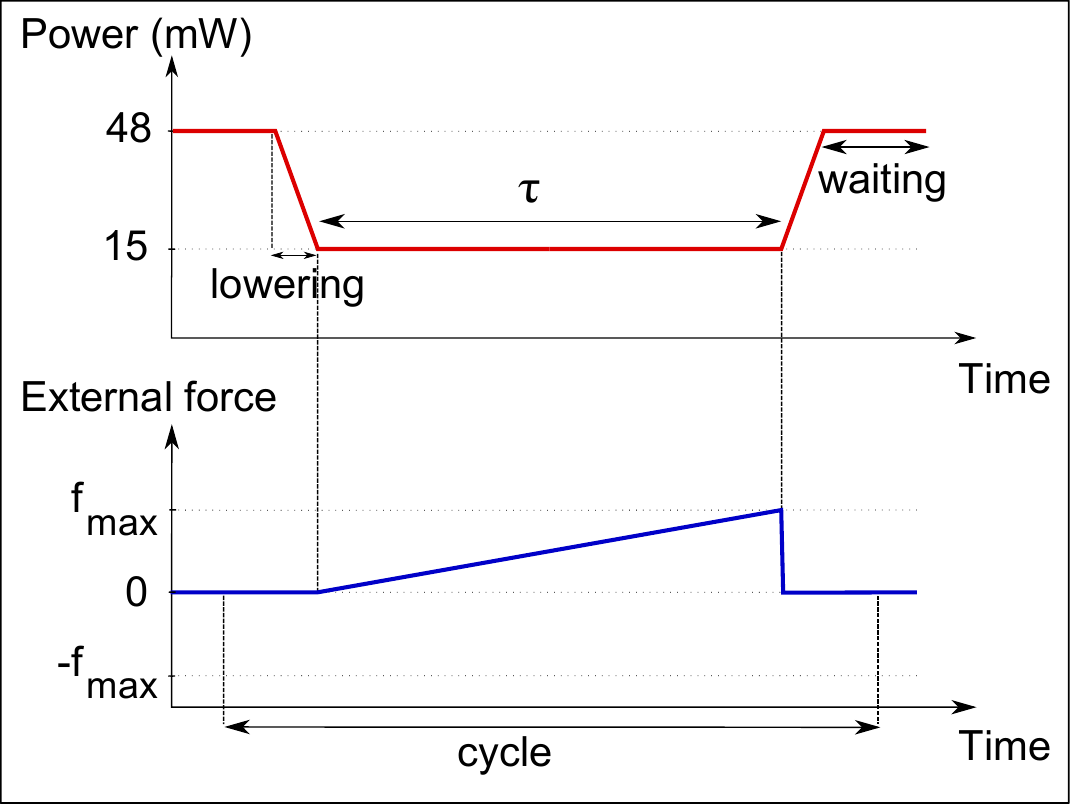}
\caption{Schematic representation of the erasure procedure.} \label{fig:schema_procedure}
\end{center}
\end{figure}

\noindent At the beginning the laser power is high ($\SI{48}{\milli\watt}$) so that the central barrier is more than $8 kT$ and the characteristic jumping time (Kramers Time) is about $\SI{3000}{\second}$, which is long compared to the time of the experiment, and the bead is trapped in one well-defined state (0 or 1). The system is left 4 $s$ with high laser power so that the bead is at equilibrium in the well where it is trapped (the relaxation time of the bead is about $\SI{0.01}{\second}$). The laser power is first lowered (in a time $T_{\text{lowering}} = \SI{1}{\second}$) to $\SI{15}{\milli\watt}$ so that the barrier is about $2.2 k_{\text{B}}T$ and the jumping time falls to $~\SI{10}{\second}$. A viscous drag force linear in time is induced by displacing the cell with respect to the laser using a closed-loop 3-axis NanoMax stage. The force is given by $f= - \gamma v$ where $\gamma=6 \pi R \eta$ ($\eta$ is the viscosity of water corrected by $3\%$ to take into account the finite thickness of the  cell) and $v$ the speed of displacement. It tilts the double-well potential so that the bead ends always in the same well (e.g. state 0) independently of where it started (see fig. \ref{fig:potential}, \textbf{c} and \textbf{d}). At the end, the force is stopped and the central barrier is raised again to its maximal value (in  a time $T_{\text{rising}} = \SI{1}{\second}$). The experimental procedure is sketched in figure \ref{fig:schema_procedure}. A procedure is fully characterized by its duration $\tau$ and the maximum value of the force applied $f_{\text{max}}$. Its efficiency is characterized by the ``proportion of success'' $P_S$, which is the proportion of trajectories where the bead ends in the chosen well (e.g. 0), independently of where it started.

\noindent Note that the position of the bead at the beginning of each procedure is actually known because the system is resetted in one state between two procedures, but this knowledge is not used by the erasure procedure (which is always the same regardless of the initial position of the bead). See \cite{ref:Nous} for more details about the experimental erasure procedure.

The ideal erasure procedure is a logically irreversible operation because the final state gives no information about the initial state. For one bit of memory \cite{ref:Landauer}, it corresponds to a change in the entropy of the system $\Delta S = -k_{\text{B}} \ln (2)$. The procedure can arbitrarily be decomposed in two kinds of sub-procedures: one where the bead starts in one well and ends in the other (e.g. $1 \rightarrow 0$) and one where the bead is initially in the same well where it should be at the end of the procedure (e.g. $0 \rightarrow 0$).
\bigskip



The two accessible quantities are $x(t)$, the position of the bead which is measured, and $f(t)$, the force which is imposed by the displacement of the cell. The derivatives are estimated using the discretization $\dot{x}(t+\Delta t/2) \approx [x(t+\Delta t) - x(t)]/\Delta t$. Starting from these quantities it is possible to measure the stochastic work $W_{\text{st}}$ done during the erasure procedure. 

For a colloidal particle confined to one spatial dimension and submitted to a conservative potential $V(x,\lambda)$, where $\lambda=\lambda(t)$ is a time-dependent external parameter, one can define the stochastic work received by the system along a single trajectory \cite{ref:reviewSeifert}:
\begin{equation}
W_{\text{st}}[x(t')]=\int_{0}^{t'} \! \frac{\partial V}{\partial \lambda} \dot{\lambda} \, \mathrm{d} t
\label{eq:stocwork}
\end{equation}

\noindent Here, since the force applied is independent of the position, the system can be described by an effective potential \cite{ref:Gaspard,ref:Bechinger,ref:SeifertBechinger} $V(x,t) = U_0(x,I(t)) - x \times f(t)$, where $U_0$ is due to the optical trapping and $I(t)$ is the intensity of the laser (see figure \ref{fig:potential}). If the bead does not jump from one well to the other during the modulation of the height of the barrier this part of the procedure does not contribute to the work received by the bead because it is done in a quasi-static way (the duration of the modulation is long compared to the relaxation time of the bead inside a single well which is about $\SI{0.01}{\second}$). Then the work can be computed only on the part of the procedure where the external force is applied (between $t=0$ and $t=\tau$) \cite{numerical_test}. When it is applied, the force is directly the control parameter, and considering that $f(t=0)=0=f(t=\tau)$, it follows that the stoc
 hastic work is equal to the classical work $W$:
\begin{equation}
W_{\text{st}}[x(\tau)]=\int_{0}^{\tau} \! -x \dot{f} \, \mathrm{d} t = \int_{0}^{\tau} \! f \dot{x} \, \mathrm{d} t = W[x(\tau)]
\end{equation}

\noindent The two integrals have been calculated for all the trajectories of all the procedures tested. Among all of them, the mean value of $|W_{\text{st}}-W|$ was about $7.10^{-4} k_{\text{B}}T$ and the the maximal difference observed was of $0.06 k_{\text{B}}T$, which is negligible.

We now analyse the results of our experiments.  
For every chosen duration $\tau$, the maximal force $f_{\text{max}}$ was set to different values (typically between \SI{10}{\femto\newton} and \SI{60}{\femto\newton}). For each set of parameters $(\tau,f_{\text{max}})$, the procedure was repeated several hundreds of times to be able to compute  statistical values. For each $\tau$, the value of $f_{\text{max}}$ is  optimized in order to be as small as possible and give a proportion of success $P_S>90\%$.  

\begin{figure}[ht!]
\begin{center}
\includegraphics[width=8cm]{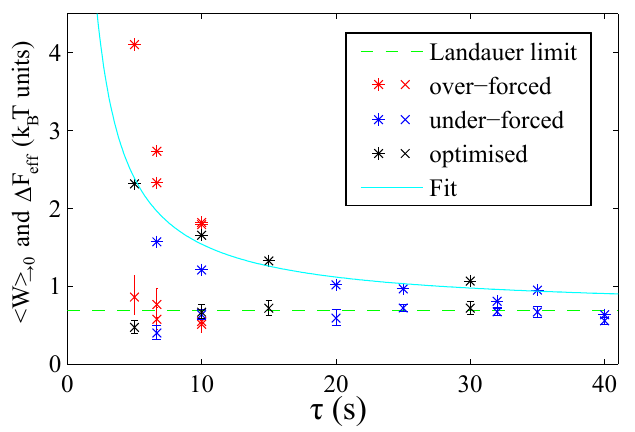}
\caption{Mean of the work ($\ast$) and effective free energy difference ($\times$) for different procedures. The over-forced procedures (red) have a proportion of success $P_S \sim 95\%$, the optimized procedures (black) have $P_S> 91\%$, the under-forced procedures have  $P_S> 83\%$ (except the last point, that has $P_S\approx 75\%$). The fit (blue line) is: $\left\langle W (\tau) \right\rangle_{\rightarrow 0}= k_{\text{B}}T \ln (2) + B/ \tau $ with $B$ a constant. For readability questions the error-bars on the mean work ($\ast$) are not shown but were estimated to be $\pm 0.15 k_{\text{B}}T$.} \label{fig:figure1}
\end{center}
\end{figure}

\noindent The trajectories where the information is erased, i.e. the ones where the bead ends where it was supposed to be (e.g. in state 0), are selected. The mean of the work received $\left\langle W \right\rangle_{\rightarrow 0}$ and the logarithm of the mean of its exponential $-\ln \left( \left\langle e^{- \beta W} \right\rangle_{\rightarrow 0} \right)$ are calculated, where $\left\langle . \right\rangle_{\rightarrow 0}$ stands for the mean on all the trajectories ending in 0. We call the value $-\ln \left( \left\langle e^{- \beta W} \right\rangle_{\rightarrow 0} \right)$ the effective free energy difference $\Delta F _{\text{eff}}$. The error bars on this value are estimated by computing the mean on the data set with $10\%$ of the points randomly excluded, and taking the maximal difference in the mean value observed by repeating this operation $1000$ times. The results are shown in figure \ref{fig:figure1}. The mean work $\left\langle W \right\rangle_{\rightarrow 0}$ dec
 reases when the duration of the procedure increases. For the optimized values of the force, it follows a law $\left\langle W (\tau) \right\rangle_{\rightarrow 0} = k_{\text{B}}T \ln (2) + B/ \tau$ where $B$ is a constant, which is the behavior for the theoretical optimal procedure \cite{ref:Gawedzki}. A least mean square fit gives $B =8.45 \, k_{\text{B}}T.s$. The effective free energy difference $\Delta F_{\text{eff}}$ is always close to the Landauer limit $k_{\text{B}}T \ln (2)$, independently of the value of the maximal force or the procedure duration.

The mean of the exponential can be computed on the sub-procedures by sorting trajectories in function of the initial position of the bead. Specifically :
\begin{equation}
\left\langle e^{- \beta W} \right\rangle_{\rightarrow 0}= \frac{M_{10}+ M_{00}}{2}
\end{equation}
\noindent where the factor 1/2 comes from the equally distributed initial state and:
\begin{equation}
M_{10}=\left\langle e^{-\beta W} \right\rangle_{1 \rightarrow 0} \quad \textrm{and} \quad M_{00}=\left\langle e^{-\beta W} \right\rangle_{0 \rightarrow 0}
\label{eq:MM}
\end{equation} 
\noindent where $\left\langle . \right\rangle_{i \rightarrow 0} $  stand for mean on the trajectories ending in 0 and starting either in $i=1$ or in $i=0$.
\noindent 

\begin{figure}[ht!]
\begin{center}
\includegraphics[width=8cm]{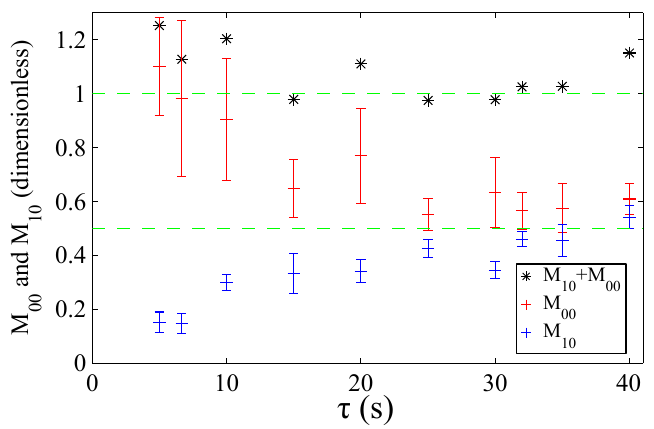}
\caption{Mean of the exponential of the work, for the sub-procedures $1 \rightarrow 0$ (blue) and $0 \rightarrow 0$ (red). For readability questions, only one value is shown for each $\tau$: it corresponds to the procedure with the highest proportion of success $P_S$.} \label{fig:figure2}
\end{center}
\end{figure}

The values $M_{10}$ and $M_{00}$ are plotted in fig. \ref{fig:figure2}. The sum $M_{10}+M_{00}$ is always close to $1$, which corresponds to the fact that $\Delta F _{\text{eff}}$ is close to $k_{\text{B}}T \ln (2)$, but $M_{00}$ decreases with $\tau$ whereas $M_{10}$ increases consequently.

\bigskip


These results can be understood in the following way:
\noindent Since the memory erasure procedure is made in a cyclic way and $\Delta S = - k_{\text{B}} \ln(2)$ it is natural to await $\Delta F = k_{\text{B}}T \ln(2)$. But the $\Delta F$ that appears in the Jarsynski equality is the difference between the free energy of the system in the initial state (which is at equilibrium) and the equilibrium state corresponding to the final value of the control parameter: $F(\lambda(\tau))-F(\lambda(0))$. Since the height of the barrier is always finite there is no change in the equilibrium free energy of the system between the beginning and the end of the procedure. Then $\Delta F = 0$, which implies $\left\langle e^{- \beta W_{\text{st}}} \right\rangle = 1$. Nevertheless Vaikuntanathan and Jarzysnki \cite{ref:Jarzynski2} have shown that when there is a difference between the actual state of the system (described by the phase-space density $\rho _{t}$), and the equilibrium state (described by $\rho ^{\text{eq}}_{t}$), the Jarzynski equality {can} be modified:
\begin{equation}
\left\langle e^{- \beta W_{\text{st}}(t)} \right\rangle _{(x,t)} = \frac{\rho ^{\text{eq}}(x,\lambda (t))}{\rho (x,t)} e^{-\beta \Delta F(t)}
\label{eq:jarzynski}
\end{equation}
\noindent Where $\left\langle . \right\rangle _{(x,t)}$ is the mean on all the trajectories that pass through $x$ at $t$.

\noindent In our procedure, the selection of  the trajectories where the information is actually erased, corresponds to fix $x$  to the chosen final well (e.g. state 0) at the time $t=\tau$. It follows that $\rho (0,\tau)$ is directly $P_{S}$, the proportion of success of the procedure, and $\rho ^{\text{eq}}(0,\lambda (\tau)) = 1/2$ \cite{sup_mat}. Then:
\begin{equation}
\left\langle e^{- \beta W(\tau)} \right\rangle _{\rightarrow 0} = \frac{1/2}{P_{S}}
\label{eq:<>0}
\end{equation}
\noindent Similarly for the trajectories that end the procedure in the wrong well (e.g. state 1) we have:
\begin{equation}
\left\langle e^{- \beta W(\tau)} \right\rangle _{\rightarrow 1} = \frac{1/2}{1-P_{S}}
\label{eq:<>1}
\end{equation}
\noindent  Taking into account  the Jensen's inequality, i.e. 
$ \left\langle e^{-x } \right\rangle \ge e^ {-\left\langle x \right\rangle}$, we  find that equations \ref{eq:<>0} and \ref{eq:<>1} imply:
\begin{equation}
\begin{array}{l}
\left\langle W \right\rangle_{\rightarrow 0} \geq k_{\text{B}} T \left[ \ln(2) + \ln(P_S) \right]  \\
\left\langle W \right\rangle_{\rightarrow 1} \geq k_{\text{B}} T \left[ \ln(2) + \ln(1-P_S) \right]
\end{array}
\end{equation}
\noindent Notice that the mean work dissipated to realize the procedure is simply: 
\begin{equation}
\left\langle W \right\rangle = P_S \times \left\langle W \right\rangle_{\rightarrow 0} + (1-P_S) \times \left\langle W \right\rangle_{\rightarrow 1}
\end{equation}
\noindent where $\left\langle . \right\rangle$ is the mean on all trajectories. Then using the previous inequalities it follows:
\begin{equation}
\left\langle W \right\rangle \geq k_{\text{B}} T \left[ \ln(2) + P_{S} \ln(P_{S}) + (1-P_{S}) \ln(1-P_{S}) \right]
\end{equation} 
\noindent which is indeed the generalization of the  Landauer's limit for $P_S < 1$. {In the limit case where  $P_S \rightarrow 1$, we have:
\begin{equation}
\left\langle e^{- \beta W} \right\rangle_{\rightarrow 0} = 1/2
\end{equation}
Since this result remains approximatively verified for proportions of success close enough to 100\%, it explains why in the experiment we find $\Delta F_{\text{eff}} \approx k_{\text{B}} T \ln(2)$. This result is not in contradiction with the classical Jarzynski equality, because if we average over all the trajectories (and not only the ones where the information is erased), we find:
\begin{equation}
\left\langle e^{- \beta W} \right\rangle = P_S \times \left\langle e^{- \beta W} \right\rangle_{\rightarrow 0} + (1-P_S) \times \left\langle e^{- \beta W} \right\rangle_{\rightarrow 1} = 1
\label{classical_jar}
\end{equation}
\noindent But it's the use of the detailed equation that allows us to find the Landauer limit. For simplicity reasons we consider $P_S = 1$ in the following part.}

To understand the evolution of $M_{10}$ and $M_{00}$, we need to consider the subprocedures $1 \rightarrow 0$ and $0 \rightarrow 0$ separately. In this case the classical Jarzynski equality does not hold because the initial conditions are not correctly tested (selecting trajectories by their initial condition introduces a bias in the initial equilibrium distribution). But Kawai and coworkers \cite{ref:VanDenBroeck} have shown that for a partition of the phase-space into non-overlapping subsets $\chi_{j}$ ($j=1,...,K$) there is a detailed Jarzynski Equality :
\begin{equation}
\left\langle e^{- \beta W} \right\rangle_{j} = \frac{\tilde{\rho}_j}{\rho_j} \left\langle e^{- \beta W} \right\rangle = \frac{\tilde{\rho}_j}{\rho_j} e^{- \beta \Delta F_{\text{eff}}}
\label{eq:vandenbroeck}
\end{equation}
with:
\begin{equation}
\rho_j = \int_{\chi_{j}} \rho(t_{a}) \, \mathrm{d}x\mathrm{d}p ~~\mbox{and}~~ \tilde{\rho}_j = \int_{\tilde{\chi}_{j}} \tilde{\rho}(t_{a}) \, \mathrm{d}x\mathrm{d}p
\end{equation}
where $\rho(t_{a})$ and $\tilde{\rho}(t_{a})$ are the phase-space densities of the system measured at the same intermediate but otherwise arbitrary point in time, in the forward and backward protocol, respectively. This type of fluctuation theorem has already been used to experimentally measure free-energy of kinetic molecular states \cite{ref:Junier,ref:Alemany}. Here, there are only two subsets $j=\{00,10\}$, defined by the position where the bead starts. By taking $t_{a}=0$ the starting point of the procedure, we have $\rho_{00}=1/2=\rho_{10}$, and $\tilde{\rho}_{00}$ (resp. $\tilde{\rho}_{10}$) identifies with the probability $\tilde{P}_{00}$ (resp. $\tilde{P}_{01}$) that the system returns into its initial state, i.e. state 0 (resp. state 1), under the time-reversed procedure. Since $e^{- \beta \Delta F _{\text{eff}}}=1/2$  it follows from eq. \ref{eq:vandenbroeck} and the definition in eq. \ref{eq:MM} that:
\begin{equation}
 M_{10} = \tilde{P}_{01} \quad \textrm{and} \quad M_{00} = \tilde{P}_{00}
 \label{eq:timereversed}
\end{equation}
\noindent 
This result is similar to the one reported in ref. \cite{ref:Ueda,ref:Toyabe} for procedures with feedback. It should be noticed that here $\tilde{P}_{01}+\tilde{P}_{00}=1$. It is reasonable to think that for time-reversed procedures (that always start in state 0) the probability of returning to state 1 is small for fast procedures and increases by increasing the duration $\tau$, which  explains qualitatively the behavior of $M_{00}$ and $M_{10}$ observed experimentally. To be more quantitative one has to measure  $\tilde{P}_{01}$ and $\tilde{P}_{00}$, but the time-reversed procedure cannot be realized experimentally, because it starts with a very fast rising of the force, which cannot be reached in our experiment.

Thus, in order to verify eq. \ref{eq:timereversed}, we performed a numerical simulation, where it is possible to realize the corresponding time-reversed procedure and to compute $\tilde{P}_{01}$ and $\tilde{P}_{00}$. Our experimental system can be described by the over-damped Langevin equation:
\begin{equation}
\gamma \dot{x} = - \frac{\partial V}{\partial x} + \xi
\end{equation}
\noindent where $\xi$ is a gaussian white noise with zero mean and correlation $\left\langle \xi(t)\xi(t') \right\rangle =2\gamma k_{\text{B}} T \delta (t-t')$.

Simple numerical simulations were made by integrating this equation with Euler method, for a set of procedures as close as possible to the experimental ones. Some results are showed in the following table:

\begin{center}
\begin{tabular}{|c|c|c|c|c|c|c|}
  \hline
  $\tau$ & $f_{\text{max}}$ & $M_{10}$ & $\tilde{P}_{01}$ & $M_{00}$ & $\tilde{P}_{00}$ & success \\
  (\si{s}) & (\si{fN}) & & & & & ($\%$) \\
  \hline
  5 & 37.7 & 0.19  & 0.19 & 0.84 & 0.81 & 97 \\
  10 & 28.3 & 0.30 & 0.30 & 0.73 & 0.70 & 96.5 \\
  20 & 18.9 & 0.45 & 0.41 & 0.63 & 0.59 & 94 \\
  30 & 18.9 & 0.45 & 0.44 & 0.60 & 0.56 & 94.5 \\
  \hline
\end{tabular}
\end{center}

The agreement between $M_{10}$ (resp. $M_{00}$) and $\tilde{P}_{01}$ (resp. $\tilde{P}_{00}$) is quantitative (the values are estimated at $\pm 0.01$), and we also retrieve the fact that $M_{10}+M_{00}$ is always close to $1$ for any set of parameters with reasonnable success rate, as in the experiments. 

\noindent It was also verified that for proportions of success \mbox{$<100\%$}, if one takes all the trajectories, and not only the ones where the bead ends in the state 0, the classical Jarzynski equality is verified: $\left\langle e^{- \beta W_{\text{st}}} \right\rangle = 1$ (for these specific simulations, $T_{\text{lowering}}$ and $T_{\text{rising}}$ were taken equal to $\SI{0.1}{\second}$ to avoid problems when the bead jumps during this phase of the procedure). This result means that the small fraction of trajectories (sometimes $<1\%$) where the bead ends the erasure procedure where it shouldn't is enough to retrieve the fact that $\Delta F = 0$.
\bigskip


As a conclusion, it has been experimentally shown that for a memory erasure procedure of a one bit system, which is a logically irreversible operation,  a detailed Jarzynski equality  allows us to retrieve the Landauer's bound for  the work done on the system independently on the speed in which the memory erasure procedure is performed. Furthermore we show that the division of the procedure into two sub-procedures is useful in order to link the work done on the system to the probability that the memory returns to its initial state under the time-reversed procedure.
These results are important  because they clarify the use of the Jarzinsky equality in irreversible operations.


\section{Acknowledgements}
We thank David Lacoste, Krzysztof Gawedzki, Luca Peliti and Christian Van den Broeck for very useful and interesting discussions.
This work has been partially supported by ESF network ``Exploring the Physics of Small Device''.

\setcounter{equation}{0}
\renewcommand{\theequation}{A.\arabic{equation}}

\section{Appendix}
\label{appendix}

\noindent Equation \ref{eq:<>0} is obtained directly if the system is considered as a two state system, but it also holds if we consider a bead that can take any position in a continuous 1D double potential along the x-axis. We place the reference $x=0$ at the center of the double potential.
\bigskip

\noindent Equation \ref{eq:jarzynski} states:
\begin{equation}
\left\langle \mathrm{e}^{- \beta W(t)} \right\rangle _{(x,t)} = \frac{\rho ^{\text{eq}}(x,\lambda (t))}{\rho (x,t)} \mathrm{e}^{-\beta \Delta F(t)}
\label{eq:paper}
\end{equation}
\noindent Where $\left\langle . \right\rangle _{(x,t)}$ is the mean on all the trajectories that pass through $x$ at $t$.
\bigskip

\noindent We choose $t=\tau$ the ending time of the procedure, and we will not anymore write the explicit dependance upon $t$ since it's always the same chosen time.

\noindent We recall that $\Delta F (\tau) = 0$ for our procedure.

\noindent We define the proportion of success, which is the probability that the bead ends its trajectory in the left half-space $x<0$:
\begin{equation}
P_S = \rho(x < 0) = \int_{-\infty}^{0} \mathrm{d}x \,  \rho (x)
\end{equation}

\noindent The conditional mean is given by:
\begin{equation}
\left\langle \mathrm{e}^{- \beta W} \right\rangle _{x} = \int \mathrm{d}W \,  \rho (W | x) \mathrm{e}^{- \beta W}
\label{eq:defmean}
\end{equation}
\noindent Where $\rho (W | x)$ is the conditional density of probability of having the value $W$ for the work, knowing that the trajectory goes through $x$ at the chosen time $\tau$.

\noindent We recall from probability properties that:
\begin{equation}
\rho (W | x) = \frac{\rho (W,x)}{\rho (x)}
\label{eq:condprob}
\end{equation}
\noindent Where $\rho (W,x)$ is the joint density of probability of the value $W$ of the work and the position $x$ through which the trajectory goes at the chosen time $\tau$.

\noindent Also: 
\begin{equation}
\rho (W | x < 0) = \frac{\int_{-\infty}^{0} \mathrm{d}x \, \rho (W,x)}{\int_{-\infty}^{0} \mathrm{d}x \, \rho (x)}=\frac{\int_{-\infty}^{0} \mathrm{d}x \, \rho (W,x)}{P_S}
\label{eq:condprobint}
\end{equation}

\bigskip

\noindent Then by multiplying equation \ref{eq:paper} by $\rho (x)$ and integrating over the left half-space $x < 0$ we have:
\begin{equation}
\int_{-\infty}^{0} \mathrm{d}x \, \rho (x) \left\langle \mathrm{e}^{- \beta W} \right\rangle _{x} = \int_{-\infty}^{0} \mathrm{d}x \, \rho ^{\text{eq}} (x)
\end{equation}

\noindent Since the double potential is symetric $\int_{-\infty}^{0} \mathrm{d}x \, \rho ^{\text{eq}} (x) = \frac{1}{2}$.

\noindent By applying definition \ref{eq:defmean} and equality \ref{eq:condprob}, it follows:
\begin{equation}
\int_{-\infty}^{0} \mathrm{d}x \, \int \mathrm{d}W \, \rho (W , x) \mathrm{e}^{- \beta W} = \frac{1}{2}
\end{equation}

\noindent Then using equality \ref{eq:condprobint}:
\begin{equation}
P_S \int \mathrm{d}W \, \rho (W | x < 0) \mathrm{e}^{- \beta W} = \frac{1}{2}
\end{equation}

\noindent finally we obtain:
\begin{equation}
\left\langle \mathrm{e}^{- \beta W} \right\rangle _{x<0} = \frac{1/2}{P_S}
\end{equation}

\noindent which is eq. \ref{eq:<>0} of the main text. This proves that eq. \ref{eq:<>0} is valid both for continous variables as shown here and for discrete variables as it is done in the main text.


\begin{thebibliography}{99}
\bibitem{numerical_test}
This condition was numerically tested with the simulations described in the last part of the article: for the parameters used, the work due to the modulation of the height of the barrier is zero on average and doesn't change for more than $5 \%$ the mean of the work $\left\langle W_{\text{st}} \right\rangle$ or the mean of its exponential $\left\langle e^{- \beta W_{\text{st}}} \right\rangle$.
\bibitem{sup_mat} 
In this way we are discretizing the space in only two subsets, but eq. \ref{eq:<>0} can be proved for a continuous distribution.
See the appendix for the detailed proof.
\end{thebibliography}
\end{document}